\documentclass[a4paper,11pt]{article}   
\usepackage{jheppub} 
\usepackage{lineno}
\usepackage[]{hyperref}
\usepackage{placeins}
\usepackage{diagbox}
\usepackage{braket}

\usepackage[utf8]{inputenc}
\usepackage{amssymb}
\usepackage{amsmath}
\usepackage{amsfonts}
\usepackage{graphicx}
\usepackage{color}
\usepackage{xspace}
\usepackage{comment}
\usepackage[normalem]{ulem}

\usepackage{afterpage}
\usepackage{float}

\usepackage{slashed}
\usepackage{ulem}
\usepackage{appendix}
\usepackage{cancel}

\usepackage{multirow,rotating}
\usepackage[dvipsnames]{xcolor}
\usepackage{orcidlink}   
\usepackage{color,bm}

\newcommand{\bea} {\begin{eqnarray}}
\newcommand{\eea} {\end{eqnarray}}

\newcommand{\beq} {\begin{equation}}
\newcommand{\eeq} {\end{equation}}

\title{\boldmath The Inert Doublet Model of Dark Matter and the LUX-ZEPLIN High-Recoil Event}

\author[a]{Lei Wang\footnote{Corresponding author}}
\author[b]{,Yang Xiao\footnote{Corresponding author}}

\affiliation[a]{Department of Physics, Yantai University, Yantai 264005, China}
\affiliation[b]{School of Physics, Henan Normal University, Xinxiang 453007, P. R. China}

\emailAdd{leiwang@ytu.edu.cn}
\emailAdd{xiaoyangphy@gmail.com}

\abstract{
The recent LUX-ZEPLIN (LZ) search reported a nuclear-recoil
event near $248\,\mathrm{keV}$. Such a high-recoil event
can be interpreted in terms of endothermic inelastic dark
matter scattering, which typically requires a mass splitting
of order a few hundred keV between the initial and final
dark states. The inert doublet model (IDM) provides a
natural realization of this scenario through the
$Z$-mediated transition $H+N\to A+N$.
In this work, we systematically examine this interpretation
in the IDM under the relevant theoretical and experimental
constraints. Our profile-likelihood analysis of the LZ
event favors a high-mass region with a best-fit point at
$m_H=1080\,\mathrm{GeV}$ and
$\delta=m_A-m_H=369\,\mathrm{keV}$ under the reference
halo assumptions. We find that the preferred mass splitting
depends on the halo velocity distribution. Including the
leading electroweak corrections to elastic scattering,
we show that dark matter captured in the Sun thermalizes
efficiently, leading to a neutrino signal that excludes
the original best-fit point. Motivated by the high speed tail associated
with the Large Magellanic Cloud (LMC), we identify a new
benchmark at $m_H=1080\,\mathrm{GeV}$ and
$\delta=635\,\mathrm{keV}$ that can explain the LZ
recoil while remaining consistent with the adopted
IceCube constraint.

}

\begin{document} 
\maketitle
\flushbottom

\section{Introduction}

The nature of dark matter (DM) remains one of the most important open questions in particle physics and cosmology. Although a wide range of astrophysical and cosmological observations provide compelling evidence for the existence of dark matter, its interactions with the Standard Model (SM) particles remain unknown. Among various dark matter candidates, weakly interacting massive particles (WIMPs) provide a particularly attractive framework, since their weak-scale interactions can naturally lead to the observed dark matter abundance through thermal freeze-out. Direct-detection experiments search for dark matter through its scattering off atomic nuclei. The LUX-ZEPLIN (LZ) experiment is currently one of the most sensitive direct-detection experiments and has placed stringent constraints on conventional spin-independent and spin-dependent dark matter-nucleon scattering cross sections.

Recently, the LZ Collaboration reported one event with a reconstructed nuclear recoil energy of $E_{\rm nr}=248\pm23~({\rm stat})\pm23~({\rm sys})~\mathrm{keV}$ \cite{LZ:2026axp}. The corresponding profile-likelihood analysis gives a global significance of approximately $2.6\sigma$, while the maximum local significance reaches approximately $3.4\sigma$. Although this excess does not constitute statistically significant evidence for dark matter, the unusually large recoil energy provides an interesting opportunity to investigate dark matter scenarios beyond the conventional elastic WIMP framework. Several particle-physics interpretations of this event have
already been proposed in Refs. \cite {Wu:2026nhi,Lou:2026idn,Freese:2026sga,Su:2026,FanReece:2026,Yamashita:2026ump,Visinelli:2026kgt,Yin:2026jnn,Unwin:2026rdp,
Jeesun:2026vzo,McCabe:2026crm,Rodd:2026tyn,Du:2026guj,Smirnov:2026aqk,Chattopadhyay:2026ryw,DiMauro:2026,Nomura:2026qyq,Bose:2026ndd,Bisal:2026khf,Bandyopadhyay:2026gjw,Das:2026uyy,Ahmed:2026qjg,Baer:2026fpy,Lee:2026jxl,Elahi:2026vlm,Dent:2026bji,deLima:2026shq,Khlopov:2008ki,Khlopov:2009hi,Roy:2026wzb}.
In particular, high-recoil events can be naturally associated with inelastic dark matter interactions. Consider a dark matter particle $\chi_1$ scattering off a nucleus and transitioning into a slightly heavier state $\chi_2$, $\chi_1+N\rightarrow\chi_2+N$ with $\delta=m_{\chi_2}-m_{\chi_1}>0$.
Such a process is referred to as endothermic inelastic scattering.

The Inert Doublet Model (IDM) provides a simple and well-motivated realization of inelastic dark matter \cite{Deshpande:1977rw,Barbieri:2006dq,LopezHonorez:2006,Arina:2009um,Bose:2024bnp}. The model extends the SM by introducing an additional scalar $SU(2)_L$ doublet, which is odd under an exact $Z_2$ symmetry, while all SM fields  are $Z_2$ even. The inert doublet does not acquire a vacuum expectation value, and its lightest neutral component is therefore stable and can serve as a dark matter candidate.
An important feature of the IDM is that the two neutral inert scalars, denoted by $H$ and $A$, have an off-diagonal coupling to the $Z$ boson. Consequently, tree-level $Z$-mediated scattering of the dark matter particle $H$ occurs through the inelastic process
$H+N\rightarrow A+N,$ provided that $m_A>m_H$.
 Therefore, the IDM provides precisely the ingredients required for an endothermic dark matter interpretation of the LZ high-recoil event. In this work, we perform a detailed profile-likelihood analysis of the LZ event with the IDM, systematically incorporating the dark matter observables and the relevant constraints. 

The paper is organized as follows. In Sec.~\ref{sec:IDM}, we introduce the inert doublet model briefly. In Sec.~\ref{sec:constraints}, we discuss the inelastic dark matter scattering process relevant to the LZ high-recoil event and summarize the theoretical, cosmological, and experimental constraints imposed on the IDM. In Sec.~\ref{sec:IDM Parameter Space}, we present the LZ profile-likelihood
results and examine their dependence on the
halo velocity distribution. In Sec.~\ref{sec:solar_constraints}, we investigate the effects of solar capture,
electroweak loop corrections, and the LMC-enhanced high-speed
tail on the viability of the LZ-favored IDM scenario
under the IceCube constraints. Finally, we present our conclusions in Sec.~\ref{sec:conclusion}.

\section{Inert Doublet Model}
\label{sec:IDM}
The Inert Doublet Model extends the SM Higgs sector by introducing an additional scalar $SU(2)_L$ doublet. We denote the SM-like Higgs doublet and the inert doublet by $H_1$ and $H_2$, respectively,
\begin{equation}
H_1=
\begin{pmatrix}
G^+\\
\dfrac{v+h+iG^0}{\sqrt{2}}
\end{pmatrix},
\qquad
H_2=
\begin{pmatrix}
H^+\\
\dfrac{H+iA}{\sqrt{2}}
\end{pmatrix}.
\end{equation}
The $H_1$ field has the electroweak vacuum expectation value (VEV) $v=246$ GeV, and $H_2$ has no VEV.

The model possesses an exact $Z_2$ symmetry under which
\begin{equation}
H_1\rightarrow H_1,
\qquad
H_2\rightarrow-H_2,
\end{equation}
while all SM fields are $Z_2$ even. 
The relevant Lagrangian is given by
\begin{equation}
{\cal L}_{\rm scalar}
=
(D_\mu H_1)^\dagger(D^\mu H_1)
+
(D_\mu H_2)^\dagger(D^\mu H_2)
-
V(H_1,H_2),
\end{equation}
where the scalar potential is
\begin{align}
V(H_1,H_2)
=&\,
\mu_1^2 H_1^\dagger H_1
+\mu_2^2 H_2^\dagger H_2
+\lambda_1(H_1^\dagger H_1)^2
+\lambda_2(H_2^\dagger H_2)^2
\nonumber\\
&+
\lambda_3(H_1^\dagger H_1)(H_2^\dagger H_2)
+\lambda_4(H_1^\dagger H_2)(H_2^\dagger H_1)
\nonumber\\
&+
\frac{\lambda_5}{2}
\left[
(H_1^\dagger H_2)^2
+(H_2^\dagger H_1)^2
\right].
\end{align}
We take all parameters in the scalar potential to be real, such that the scalar sector is CP conserving.

The parameter $\mu_1^2$ is fixed by the scalar potential minimum conditions
\beq \label{sminconds}
\mu_1^2=- \lambda_1 v^2.
\eeq
After electroweak symmetry breaking, the mass of the SM-like Higgs boson ($h$) is
\begin{equation}
m_h^2=2\lambda_1v^2,
\end{equation}
while the masses of the $Z_2$-odd scalar particles are
\begin{equation}
m_{H^\pm}^2
=
\mu_2^2+\frac{\lambda_3v^2}{2},~~
m_H^2 = m_A^2 + \lambda_5 v^2,~~
m_A^2
=m_{H^\pm}^2 + \frac{1}{2}(\lambda_4-\lambda_5)v^2.
\end{equation}
At the one-loop level, the charged and neutral inert scalars acquire a radiatively induced mass splitting of several hundred MeV \cite{Cirelli:2005uq}. In contrast, the two neutral states remain nearly degenerate, with their mass difference controlled by the $\lambda_5$ coupling.
For $H$ to be the lightest $Z_2$-odd particle and hence a viable dark matter candidate, we require
$\lambda_5<0$.  It is convenient to define $\lambda_L=
\frac{\lambda_3+\lambda_4+\lambda_5}{2}$, which controls the interaction term between a pair of $H$ and the SM-like Higgs boson.

The small mass splitting between the neutral inert scalars is protected by an approximate global symmetry of the IDM. In the limit $\lambda_5\to 0$, the Lagrangian acquires a global $U(1)$ symmetry under $H_2\to e^{i\alpha}H_2$. Since the inert doublet has no Yukawa interactions and does not develop a vacuum expectation value, this symmetry remains unbroken. The two neutral states are then exactly degenerate, and radiative corrections cannot generate a mass splitting in this limit. For $\delta=m_A-m_H\ll m_H$, the corresponding coupling is
\begin{equation}
\lambda_5=-\frac{m_A^2-m_H^2}{v^2}
\simeq-\frac{2m_H\delta}{v^2}
\simeq-1.3\times10^{-5}.
\end{equation}
Thus, the small mass splitting is stable against radiative corrections, provided no additional interactions explicitly break the inert-doublet $U(1)$ symmetry.

The interactions between a gauge boson and two scalar fields are obtained from the gauge-kinetic Lagrangian,
\begin{align} \label{eq:SSV}
\mathcal{L}_{SSV} = &\frac{g}{2}W^+_\mu\left((H^-\overset{\leftrightarrow}{\partial}^\mu A)+i(H^-\overset{\leftrightarrow}{\partial}^\mu H)+h.c. \right)\notag\\
&-\frac{g}{2c_W}Z_\mu\left(A\overset{\leftrightarrow}{\partial}^\mu H\right)
+\left(ie\gamma_\mu+i\frac{g(c_W^2-s_W^2)}{2c_W}Z_\mu \right)(H^\mp\overset{\leftrightarrow}{\partial}^\mu H^\pm),
\end{align}
where $c_W = \cos\theta_W$ and $s_W = \sin\theta_W$ with $\theta_W$ denoting the Weinberg angle. 

The SM fermions acquire masses through Yukawa interactions with $H_1$,
\beq \label{yukawacoupling} 
- {\cal L} = y_u\,\overline{Q}_L \,
\tilde{{ H}}_1 \,u_R +\,y_d\,
\overline{Q}_L\, {H}_1 \, d_R  \, + \, y_l\,\overline{L}_L \, {H}_1
\,e_R \,+\, \mbox{h.c.}\,, 
\eeq
where $y_u$, $y_d$ and $y_\ell$ are $3 \times 3$ matrices in family space. 
Due to the exact $Z_2$ symmetry, the inert field $H_2$ has no Yukawa interactions 
with fermions.

\section{Inelastic Scattering at LZ and Relevant Constraints}
\label{sec:constraints}
Having introduced the scalar spectrum and the relevant interactions of
the inert doublet model, we now turn to its implications for the LZ
high-recoil event. The analysis is divided into two parts. In the first subsection, we calculate the nuclear-recoil spectrum induced by the $Z$-mediated inelastic transition and construct the corresponding LZ likelihood. The theoretical, cosmological, and complementary
experimental constraints imposed on the model parameters are discussed
separately in Sec.~\ref{sec:other_constraints}.

\subsection{Inelastic scattering}
\label{sec:inelastic}

The off-diagonal $ZHA$ interaction introduced in Sec.~\ref{sec:IDM} provides a
characteristic direct-detection channel of the IDM. Since the $Z$ boson
connects the two neutral inert scalars rather than coupling diagonally to
$H$, the corresponding nuclear scattering process is $HN\to AN$,
where $N$ denotes a target nucleus. For $m_A>m_H$, the final dark-sector
state is heavier than the incoming one, so that the scattering is
endothermic. This realizes the standard inelastic dark matter mechanism
within the IDM and is particularly
relevant to the recent LZ high-recoil event~\cite{LZ:2026axp}.

The kinematics of the process is controlled by the neutral-scalar mass
splitting
\begin{equation}
    \delta \equiv m_A-m_H>0 .
    \label{eq:delta}
\end{equation}
For a nuclear recoil energy $E_R$, energy and momentum conservation require
the incoming dark matter velocity to satisfy
\begin{equation}
    v_{\rm min}^{N}(E_R)
    =
    \frac{1}{\sqrt{2M_N E_R}}
    \left(
        \frac{M_N E_R}{\mu_{HN}}+\delta
    \right),
    \label{eq:vmin}
\end{equation}
where $M_N$ is the nuclear mass and
$\mu_{HN}=m_HM_N/(m_H+M_N)$ is the reduced mass. The effect of the mass
splitting can be seen more directly by minimizing $v_{\rm min}^{N}$ with
respect to the recoil energy, which gives
\begin{equation}
    E_R^\star=\frac{\mu_{HN}}{M_N}\delta ,
    \qquad
    v_{\rm thr}^{N}
    =
    \sqrt{\frac{2\delta}{\mu_{HN}}}.
    \label{eq:vthreshold}
\end{equation}
The endothermic transition therefore introduces a nonzero velocity
threshold $ v_{\rm thr}^{N}$, below which the transition is kinematically forbidden. As the mass
splitting increases, progressively faster dark matter particles are
required to overcome the energy cost of producing the heavier state
$A$. Once $v_{\rm thr}^{N}$ approaches the upper end of the Galactic
velocity distribution, only the high-velocity tail of the dark matter
population can contribute to the scattering rate.

At the same time, the recoil energy $E_R^\star$ moves to larger values with increasing $\delta$. Recoils far below
$E_R^\star$ require rapidly increasing incoming velocities and can
therefore become kinematically inaccessible before the high-recoil
region does. Consequently, a mass splitting of a few hundred keV
simultaneously suppresses the conventional low-energy recoil spectrum
and concentrates the surviving events in a narrow high-recoil window.
This provides a simple kinematic reason why endothermic scattering is
particularly well suited to the recent LZ high-recoil event
~\cite{Su:2026,FanReece:2026,DiMauro:2026}.

This same kinematic feature also makes the predicted rate sensitive to the
fastest dark matter particles in the Galactic halo. We adopt the Standard
Halo Model with
\begin{equation}
    \rho_0=0.3~{\rm GeV\,cm^{-3}},
    \qquad
    v_0=238~{\rm km\,s^{-1}},
    \qquad
    v_{\rm esc}=544~{\rm km\,s^{-1}},
    \label{eq:halo_parameters}
\end{equation}
and take $v_{\rm lab}=254~{\rm km\,s^{-1}}$. For the mass splittings of
interest, $v_{\rm min}^N$ can approach the upper end of the laboratory-frame
velocity distribution. The inferred signal region therefore depends more
strongly on the high-velocity tail than in conventional elastic WIMP
scattering~\cite{FanReece:2026,McCabe:2026crm}.

Once the kinematically accessible region is specified, the normalization
of the IDM signal is essentially fixed by the electroweak interaction.
The relevant coupling is given in Eq. (\ref{eq:SSV}),
which leads to coherent $Z$-mediated scattering on nuclei. At zero momentum
transfer, the corresponding cross section can be written as
\begin{equation}
    \sigma_N^{0}
    =
    \frac{G_F^2\mu_{HN}^2}{2\pi}
    Q_{W,N}^{\,2},
    \qquad
    Q_{W,N}
    =
    N_N-\left(1-4\sin^2\theta_W\right)Z_N ,
    \label{eq:sigma_inel_zero}
\end{equation}
where $Z_N$ and $N_N$ are the proton and neutron numbers of isotope $N$.
Unlike a generic phenomenological inelastic dark matter model, the
scattering strength is therefore not an additional parameter that can be
adjusted independently of the particle model. Once $m_H$ and $\delta$ are
specified, the dominant interaction strength follows from the electroweak
coupling.

For the recoil energies relevant here, the momentum transfer is large
enough that the finite-size nuclear response must be retained. We therefore
write
\begin{equation}
    \frac{d\sigma_N}{dE_R}
    =
    \frac{M_N\sigma_N^{0}}
         {2\mu_{HN}^{2}v^{2}}
    F_{W,N}^{2}(q),
    \qquad
    q=\sqrt{2M_N E_R},
    \label{eq:dsigma_inel}
\end{equation}
where $F_{W,N}^{2}(q)$ denotes the finite momentum coherent weak nuclear response and we numerically evaluate this using
{\tt WimPyDD}~\cite{WimPyDD}. The differential recoil rate is then obtained
by summing over the naturally occurring xenon isotopes,
\begin{equation}
    \frac{dR_{\rm inel}}{dE_R}
    =
    \frac{\rho_H}{m_H}
    \sum_N
    \frac{X_N}{M_N}
    \int_{v>v_{\rm min}^{N}(E_R)}
    d^3v\,
    f_{\rm lab}(\mathbf v)\,
    v\,
    \frac{d\sigma_N}{dE_R},
    \label{eq:rate_inel}
\end{equation}
where $X_N$ is the isotope mass fraction, $f_{\rm lab}(\mathbf v)$ is the
dark matter velocity distribution in the laboratory frame, and $\rho_H$
denotes the local density of the $H$ component. In the parameter region
where $H$ accounts for the observed dark matter abundance, we take
$\rho_H=\rho_0$.

To compare this spectrum with the LZ observation, we fold the theoretical
recoil rate with the detector efficiency and energy response,
\begin{equation}
    \frac{dN_s}{dE_{\rm obs}}
    =
    {\cal E}
    \int dE_R\,
    \epsilon(E_R)
    G(E_{\rm obs},E_R)
    \frac{dR}{dE_R},
    \label{eq:detector_response}
\end{equation}
where ${\cal E}=2.84~{\rm tonne\,yr}$ is the exposure,
$\epsilon(E_R)$ is the nuclear-recoil efficiency, and
$G(E_{\rm obs},E_R)$ describes the reconstructed-energy response. For completeness, the Higgs-mediated elastic contribution calculated
with \texttt{micrOMEGAs-7.1.4} \cite{Belanger:2001fz,Alguero:2023zol,Belanger:2026asz} is also included in the total recoil spectrum. This contribution is numerically negligible in the high-recoil region, so that the profile likelihood is determined almost entirely by the $Z$-mediated inelastic transition. We then construct an extended likelihood using the publicly available LZ high-recoil information.

For the observed event with reconstructed nuclear recoil energy
$E_{\rm nr}=248\pm23\,({\rm stat})\pm23\,({\rm sys})~\mathrm{keV}$,
we use an extended one-event likelihood in reconstructed recoil energy.
The likelihood incorporates both the expected number of accepted events
and their energy distribution. For the elastic and inelastic recoil
channels, we calculate the corresponding accepted signal yields as
\begin{equation}
    N_{s,i} = {\cal E}\int dE_R\,\epsilon(E_R)
    \frac{dR_i}{dE_R},
    \qquad
    N_s = N_{s,\mathrm{el}} + N_{s,\mathrm{inel}},
    \label{eq:signal_yield}
\end{equation}
where $i$ labels the elastic and inelastic contributions.
The corresponding reconstructed-energy spectra are obtained
by applying the detector response in
Eq.~\eqref{eq:detector_response}.
We combine the two quoted energy uncertainties in quadrature,
giving a Gaussian width of $32.5\,\mathrm{keV}$.
After normalizing each folded spectrum over the reconstructed-energy
range $50$--$300\,\mathrm{keV}$, we weight the resulting densities
by their accepted yields to construct the signal energy
distribution $p_s(E)$.

For the background, we assume a flat energy distribution,
$p_b(E)=1/(250\,\mathrm{keV})$, and constrain its expected
yield $N_b$ using the LZ estimate of $0.0106\pm0.0008$
events in the high-$S1_c$ interval containing the
candidate~\cite{LZ:2026axp}.
The resulting likelihood is
\begin{equation}
  {\cal L}
  =e^{-(a_\rho N_s+N_b)}
   \left[
     a_\rho N_s\frac{p_s(E_{\rm nr}/a_E)}{a_E}
     +N_b p_b(E_{\rm nr})
   \right]\pi(\boldsymbol{\eta}),
  \label{eq:LZ_likelihood}
\end{equation}
where $a_\rho=e^{0.15z_\rho}$ accounts for uncertainty in the local
dark matter density, while $a_E=e^{0.03z_E}$ allows a residual
energy-scale variation in the signal distribution at fixed accepted
yield. The factor $\pi(\boldsymbol{\eta})$ contains the Gaussian
constraint on $N_b$ and standard-normal constraints on $z_\rho$
and $z_E$, with $\boldsymbol{\eta}=(N_b,z_\rho,z_E)$.
At each IDM parameter point, $N_s$ is the nominal signal yield
predicted by the model.

To obtain the profile in the $(m_H,\delta)$ plane, we minimize
the negative log likelihood over the remaining IDM parameters
in the retained sample and over the nuisance parameters
$\boldsymbol{\eta}$,
\begin{align}
  {\rm NLL}_{\rm prof}(m_H,\delta)
    &= \min_{\Delta M_\pm,\lambda_L,\lambda_2,\boldsymbol{\eta}}
       [-\ln{\cal L}],
       \label{eq:profileNLL}\\
  q(m_H,\delta)
    &=2\bigl[{\rm NLL}_{\rm prof}(m_H,\delta)
             -{\rm NLL}_{\rm min}\bigr].
       \label{eq:qstat}
\end{align}
We use $q=2.30$ and $6.18$ to define the nominal two-parameter
$1\sigma$ and $2\sigma$ profile regions.
Appendix~\ref{app:lz_validation} compares the calculated recoil
spectra with the public LZ benchmarks and tests how the preferred
splitting depends on the background and energy-response
prescription.


\subsection{Relevant theoretical and other experimental constraints}
\label{sec:other_constraints}

The scalar potential must be bounded from below in all
directions of field space. At tree level, this requirement
leads to the following conditions~\cite{LopezHonorez:2006}:
\begin{equation}
\begin{gathered}
\lambda_1>0,\qquad \lambda_2>0,\\
\lambda_3+2\sqrt{\lambda_1\lambda_2}>0,\\
\lambda_3+\lambda_4-|\lambda_5|
+2\sqrt{\lambda_1\lambda_2}>0.
\end{gathered}
\end{equation}

While these conditions ensure that the potential is stable
at large field values, the electroweak vacuum must also
correspond to its global minimum. In the IDM, preserving
the $Z_2$ symmetry requires an inert vacuum with
$\langle H_2\rangle=0$. We further impose
$\lambda_4-|\lambda_5|<0$. This condition makes the lightest neutral inert scalar lighter than the charged inert scalar and rules out a charge-breaking vacuum \cite{Belyaev:2016lok}.
The inert vacuum must also be energetically preferred
over other possible stationary configurations.
In particular, when an inert-like stationary point
exists ($\mu_2^2<0$), the inert vacuum lies below it
provided that \cite{Belyaev:2016lok}
\begin{equation}
  \frac{\mu_1^2}{\sqrt{\lambda_1}}
  <
  \frac{\mu_2^2}{\sqrt{\lambda_2}}.
  \label{eq:global_inert}
\end{equation}
Accordingly, we restrict the parameter space
to configurations in which the inert vacuum
is the global minimum of the tree-level potential.

In addition to these vacuum conditions, the quartic
scalar interactions are constrained by perturbative
unitarity. The eigenvalues of the scalar scattering
matrix must satisfy~\cite{Kanemura:1993hm,Akeroyd:2000wc}

\begin{equation}
\label{unitarity}
|a_\pm|,\ |b_\pm|,\ |c_\pm|,\
|{\tt e}_\pm|,\ |{\tt f}_\pm|,\ |{\tt g}_\pm|
\leq 8\pi,
\end{equation}
where
\begin{align}
a_\pm &= 3(\lambda_1+\lambda_2)
\pm\sqrt{9(\lambda_1-\lambda_2)^2
+(2\lambda_3+\lambda_4)^2},
\nonumber\\
b_\pm &= (\lambda_1+\lambda_2)
\pm\sqrt{(\lambda_1-\lambda_2)^2+\lambda_4^2},
\nonumber\\
c_\pm &= (\lambda_1+\lambda_2)
\pm\sqrt{(\lambda_1-\lambda_2)^2+\lambda_5^2},
\nonumber\\
{\tt e}_\pm &= \lambda_3+2\lambda_4\pm3\lambda_5,
\nonumber\\
{\tt f}_\pm &= \lambda_3\pm\lambda_4,
\nonumber\\
{\tt g}_\pm &= \lambda_3\pm\lambda_5.
\end{align}

The additional scalar particles in the IDM contribute to the electroweak gauge-boson vacuum polarization functions and are therefore constrained by electroweak precision measurements. These effects are conventionally described by the oblique parameters $S$ and $T$ \cite{He:2001tp,Haber:2010bw},
\begin{eqnarray} 
 S\; &=& \; \frac{1}{\pi M_Z^2}\,  \Biggl[  
 \mathcal{B}_{22}(M_Z^2;M_H^2,M_A^2)  -\; \mathcal{B}_{22}(M_Z^2;{M^2_{H^\pm}},{M^2_{H^\pm}})
\Biggr]\, \nonumber,\\ 
T\;  &=&\;  \frac{1}{16\pi M_W^2s_W^2}\,  \Biggl[ 
 \mathcal{F}(M_{H^\pm}^2,M_H^2) - \mathcal{F}(M_H^2,M_A^2) +\; \mathcal{F}(M_{H^\pm}^2,M_A^2)  \Biggr]\,  .
\end{eqnarray} 
The loop functions are given by
\bea
\label{eqnb}
\mathcal{B}_{22}(q^2;m_1^2,m_2^2) &\equiv&
B_{22}(q^2;m_1^2,m_2^2)-B_{22}(0;m_1^2,m_2^2),\nonumber\\
\mathcal{F}(m_1^2,m_2^2) &=& \frac{1}{2}\, (m_1^2+m_2^2)-\frac{m_1^2m_2^2}{m_1^2-m_2^2}\;
\log{\left(\frac{m_1^2}{m_2^2}\right)}, 
\eea
where
\begin{equation}
B_{22}(q^2;m_1^2,m_2^2) = \frac{1}{4}\,  [m_1^2+m_2^2-\frac{1}{3}\, q^2]-\frac{1}{2}\,\int^1_0 dx\; X\; \log{(X-i\epsilon)}, 
\end{equation}
and $X \;\equiv\; m_1^2\, x + m_2^2\, (1-x) -q^2\, x(1-x)$.

To ensure consistency with experimental data, parameter points are required to lie within the 
$2\sigma$ confidence region for both the $S$ and $T$ parameters. 
This corresponds to $\chi^2<6.18$ for two degrees of freedom. 
The corresponding fit results can be found in Ref. \cite{ParticleDataGroup:2020ssz},
\beq
S=0.00\pm 0.07,~~  T=0.05\pm 0.06, 
\eeq
with a correlation coefficient $\rho_{ST}$ = 0.92.
Constraints from the oblique parameters favor a mass spectrum in which either $H$ or $A$ is nearly degenerate with the charged Higgs boson.

In the model, the tree-level couplings of the SM-like Higgs boson to fermions and gauge bosons are identical to those in the SM,
 which is consistent with the current measurements of the 125 GeV Higgs boson. 
At the LHC, the inert scalars are predominantly produced through electroweak pair-production processes, 
leading to final states with leptons accompanied by missing transverse momentum.
 The CMS Collaboration has performed a dedicated search for the pair production of new scalars predicted by the IDM. 
The observed exclusion reaches $m_H=108$ GeV for a mass splitting of $m_A-m_H=78$ GeV \cite{CMS:2026cjr}. As the mass splitting decreases,
the leptons in the final state become increasingly soft, making the signal more challenging to probe at the LHC.

The lightest inert scalar $H$ is stable due to the exact $Z_2$ symmetry and can constitute dark matter. Its thermal relic abundance is determined by annihilation processes $HH\rightarrow {\rm SM\,SM}$, and co-annihilation processes $HA,HH^{\pm},AA,H^{\pm}H^{\mp},AH^{\pm} \rightarrow {\rm SM\,SM}$ when the mass splittings among the inert scalars are sufficiently small.
The viable dark matter mass regimes in the IDM have been extensively investigated in the literature; 
see, for example, Refs.~\cite{Deshpande:1977rw,Barbieri:2006dq,LopezHonorez:2006,Hambye:2009pw,Kalinowski:2020rmb}. 
In the resonance region of $m_H\simeq \frac{m_h}{2}$, 
dark matter annihilation is enhanced by the Higgs resonance, with $b\bar{b}$ and $W^+W^-$ being the dominant final states.
 In this regime, the observed relic abundance can be accommodated while remaining consistent with the relevant experimental constraints.
For intermediate masses, approximately $160~\mathrm{GeV}\lesssim m_H\lesssim500~\mathrm{GeV}$, 
the gauge-mediated annihilation process $HH\to W^+W^-$ is sufficiently efficient, and the resulting relic abundance is typically below the observed value.
 In the high-mass regime, $m_H>$ 500 GeV, 
achieving the observed dark matter abundance generally requires small mass splittings among the three inert 
scalars, roughly $\leq$ 10 GeV. 

We require the thermal relic
abundance to be consistent with the observed value reported by the Planck \cite{Planck:2015bpv}
\begin{equation}
\Omega_{\rm DM}h^2 = 0.1198\pm 0.0012.
\end{equation}

The dark matter candidate $H$ can scatter elastically off nuclei through Higgs exchange,
$H+N\rightarrow H+N$,
with a spin-independent cross section controlled by $\lambda_L$. 
We require the theoretical value to satisfy the latest
exclusion limits from the LZ experiment \cite{LZ:2022lsv}.

Indirect searches for dark matter in dwarf spheroidal galaxies can constrain its annihilation properties. We adopt the combined constraints summarized in Ref.~\cite{Fermi-LAT:2025gei}, which incorporate the results from Fermi-LAT~\cite{Fermi-LAT:2015att}, HAWC~\cite{HAWC:2017mfa}, HESS~\cite{HESS:2018kom}, MAGIC~\cite{MAGIC:2021mog}, and VERITAS~\cite{VERITAS:2017tif}.
The package \texttt{micrOMEGAs-7.1.4} \cite{Belanger:2001fz,Alguero:2023zol,Belanger:2026asz} is employed to calculate the present-day thermally averaged tree level annihilation cross-section, the relic density, spin-independent cross section.


\section{IDM Parameter Space and the LZ Fit}
\label{sec:IDM Parameter Space}

\begin{figure*}[t]
    \centering

    \includegraphics[width=0.485\textwidth]
    {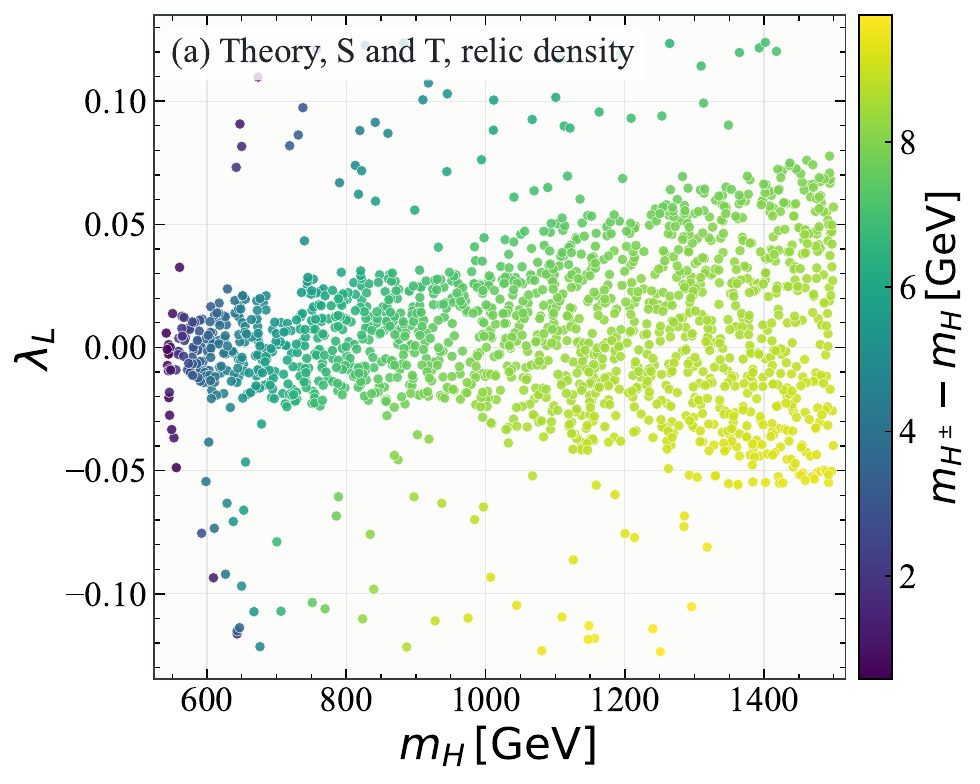}
    \hfill
    \includegraphics[width=0.485\textwidth]
    {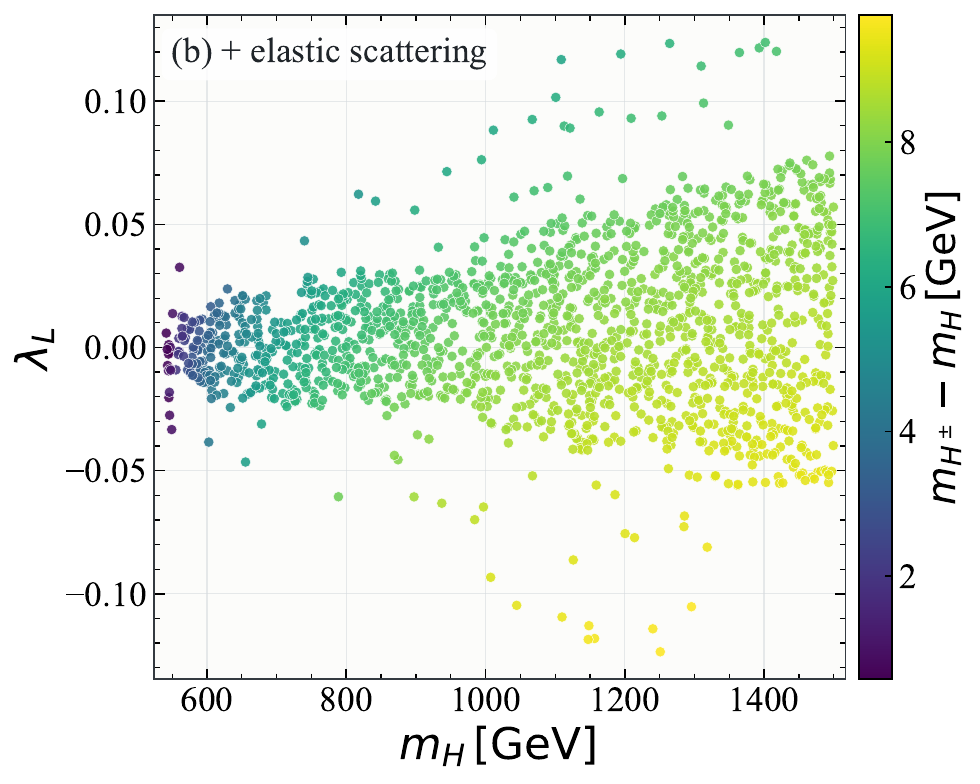}

    \par\vspace{0.6em}

    \includegraphics[width=0.485\textwidth]
    {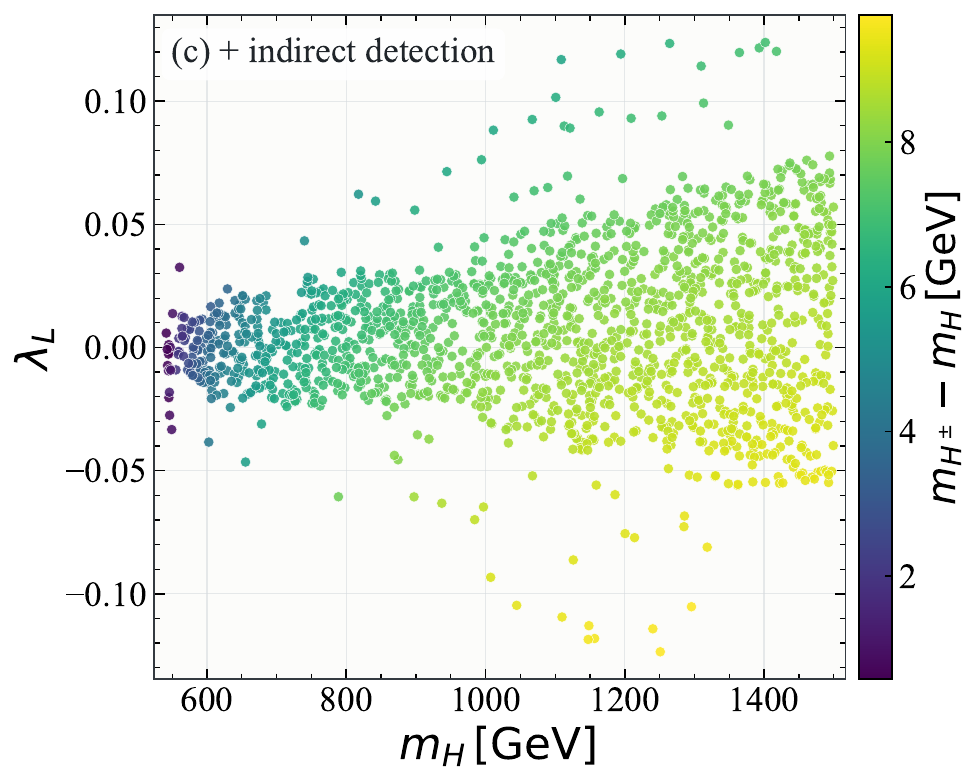}
    \hfill
    \includegraphics[width=0.485\textwidth]
    {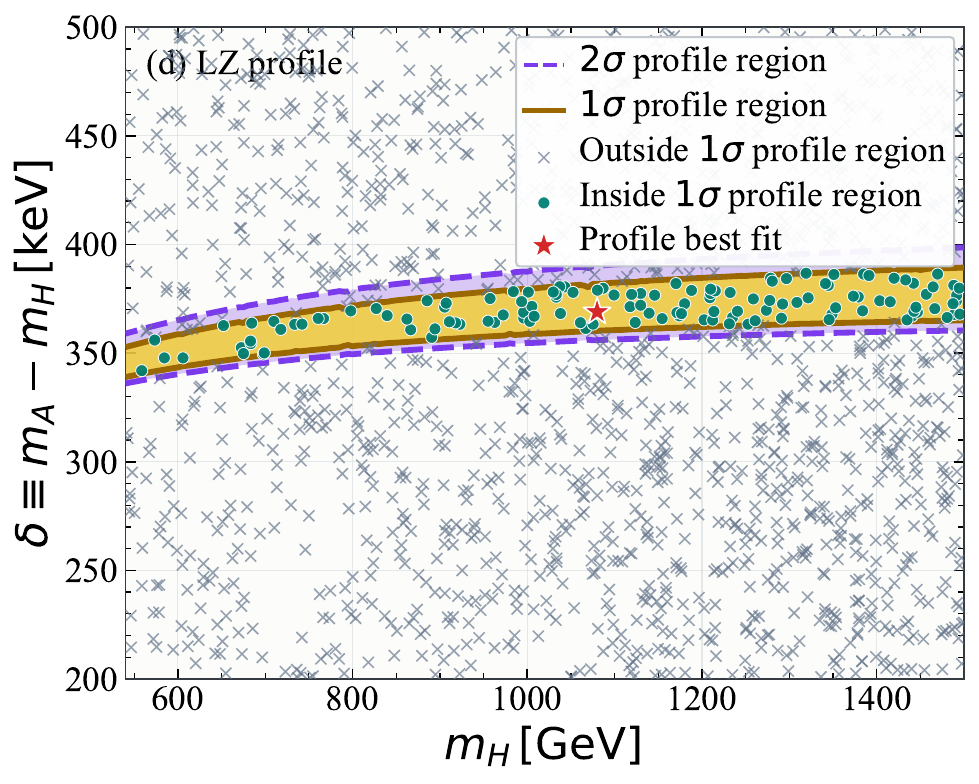}

    \caption{%
        Successive constraints on the IDM parameter space.
        The upper left panel shows the parameter points satisfying the
        theoretical consistency, the $S$ and $T$ parameters and relic density constraints.
        The upper right panel additionally imposes the elastic-scattering
        constraint.
        The lower left panel further includes the indirect-detection
        constraints.
        The lower right panel presents the corresponding LZ
        profile likelihood regions after imposing all preceding constraints.
    }
    \label{fig:idm_constraints}

\end{figure*}

We consider the following scalar mass spectrum:
\begin{align}
500~ {\rm GeV} \leq m_H \leq 1.5~ {\rm TeV},~~&  200~ {\rm keV} \leq \delta \leq 500~ {\rm keV},\nonumber\\~0.5 ~{\rm GeV} \leq m_{H^\pm}&-m_H\leq 10~ {\rm GeV}.
\end{align}
As discussed above, the CMS constraints on IDM scalar pair production are automatically satisfied throughout the mass spectrum considered here and thus require no further consideration in our scan. We carefully check all remaining theoretical and experimental constraints including theoretical consistency, electroweak precision observables the $S$ and $T$ parameters, the relic density, direct-detection constraints from elastic scattering, and indirect-detection limits.

Figure~\ref{fig:idm_constraints} summarizes the viable parameter space
after imposing the theoretical and experimental constraints discussed in
Sec.~\ref{sec:constraints}. 
In the high-mass region, co-annihilation processes, in addition to dark matter pair annihilation, play an important role in obtaining the correct relic density. As shown in the upper-left panel, for $500~\mathrm{GeV} \leq m_H \leq 1500~\mathrm{GeV}$, the observed relic density can be reproduced when the mass splitting between $m_H$ and $m_A$ is of the order of several hundred keV, while the splitting between $m_{H^\pm}$ and $m_H$ is of the order of several GeV.

The elastic direct-detection constraint further reduces the parameter
space with relatively large $|\lambda_L|$, while leaving the compressed
high-mass region. This is particularly important for the present
interpretation because the conventional elastic signal and the
high-recoil signal are controlled by different interactions. The
Higgs-mediated process $HN\to HN$ is governed by $\lambda_L$, whereas the
inelastic transition $HN\to AN$ arises from the electroweak $ZHA$
coupling. The small values of $\lambda_L$ favored by elastic
direct-detection limits therefore do not suppress the inelastic channel
of interest. We also impose the limits from indirect searches for dark matter in dwarf spheroidal galaxies. In the parameter region surviving the
relic density and elastic-scattering constraints, these bounds hardly
exclude any additional points, as can be seen by comparing the upper
right and lower left panels.

We next examine the LZ high-recoil likelihood using the surviving
parameter points. The lower right panel of
Fig.~\ref{fig:idm_constraints} shows a narrow preferred band in the
$(m_H,\delta)$ plane. The profile best-fit point is
\begin{equation}
\begin{split}
m_H &=1080~{\rm GeV}, \qquad
\delta =369~{\rm keV},\\
m_{H^\pm}-m_H &=8.17~{\rm GeV}, \qquad
\lambda_L=-1.92\times10^{-4},
\end{split}
\label{eq:bestfit}
\end{equation}
with $\lambda_2=4.06$ and
$\Omega_H h^2\simeq0.12$. The best-fit point therefore lies naturally
within the compressed high-mass region selected by the relic-density
requirement and simultaneously favors a very small Higgs-portal
coupling.

A notable feature of the likelihood is that the neutral-scalar mass
splitting is much more tightly selected than the overall dark matter
mass. This behavior follows from the endothermic kinematics discussed in
Sec.~\ref{sec:inelastic}. Increasing $\delta$ raises the energy cost of
the $H\to A$ transition and rapidly pushes the required incoming velocity
toward the upper end of the Galactic dark matter distribution. Once the
splitting becomes too large, the population of particles capable of
producing the observed recoil is strongly depleted. For a smaller
splitting, lower recoil energies become increasingly accessible and the
signal is less concentrated in the high-recoil region. The observed
event therefore selects a relatively narrow interval of $\delta$. 

\subsection{Dependence on the Galactic inputs}
\label{sec:Galactic inputs}

\begin{figure}[htbp]
\centering
\includegraphics[width=.9\linewidth]{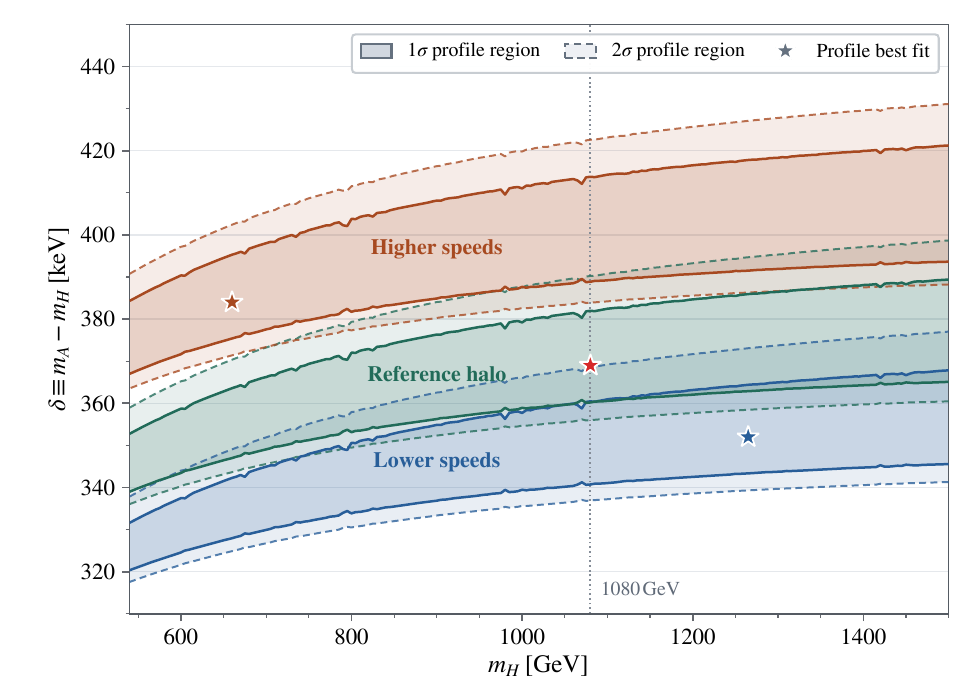}
\caption{
LZ profile-likelihood regions for three halo assumptions,
subject to the same constraints as in the lower-right panel
of Fig.~\ref{fig:idm_constraints}.
The solid and dashed contours correspond to the nominal
$1\sigma$ and $2\sigma$ regions
($q=2.30$ and $6.18$), respectively.
The stars indicate the best-fit points for each halo,
while the dotted vertical line marks
$m_H=1080\,\mathrm{GeV}$.}
\label{fig:halo_profile_comparison}
\end{figure}

To examine how the assumed halo velocity distribution affects
the IDM interpretation of the LZ events,
Fig.~\ref{fig:halo_profile_comparison} compares the
profile-likelihood regions obtained under three halo
assumptions. Starting from the reference values
$(v_0,v_{\rm esc})=(238,544)\,\mathrm{km\,s^{-1}}$,
we vary the characteristic halo speed $v_0$ and the
Galactic escape speed $v_{\rm esc}$ together, while
keeping the same IDM sample, local dark matter density
$\rho_0=0.3\,\mathrm{GeV\,cm^{-3}}$, and laboratory
speed $v_{\rm lab}=254\,\mathrm{km\,s^{-1}}$.
The green region reproduces the reference result in the
lower-right panel of Fig.~\ref{fig:idm_constraints},
with the red star indicating the best-fit point given
in Eq.~\eqref{eq:bestfit}.

The comparison shows that the lower-speed halo (blue)
favors smaller mass splittings, whereas the higher-speed
halo (orange) favors larger ones.
This behavior can be understood from the kinematics of
inelastic scattering. At a given recoil energy, a larger
mass splitting requires a higher minimum incident dark
matter speed. Increasing $v_0$ enhances the population
of fast-moving particles, while increasing $v_{\rm esc}$
extends the high-speed tail of the distribution.
Consequently, larger mass splittings can still produce
a comparable rate of high-energy recoil events.

The halo assumptions also affect the preferred dark
matter mass, as reflected by the positions of the
best-fit stars. However, since the likelihood varies
only weakly along each elongated region, we also
compare the preferred mass splittings at a fixed
dark matter mass. Table~\ref{tab:halo_profile_comparison}
summarizes the global best-fit points and the results
at $m_H=1080\,\mathrm{GeV}$, indicated by the dotted
line in Fig.~\ref{fig:halo_profile_comparison}.
At this mass, the preferred splitting changes from
$369\,\mathrm{keV}$ for the reference halo to
$349\,\mathrm{keV}$ and $400\,\mathrm{keV}$ for the
lower- and higher-speed halos, respectively.
These shifts of $20$--$31\,\mathrm{keV}$ are
comparable to or larger than the $20$--$25\,\mathrm{keV}$
widths of the nominal $1\sigma$ slices.
The inferred mass splitting is therefore sensitive
to the assumed halo velocity distribution,
particularly its high-speed tail.

\begin{table}[htbp]
\centering
\setlength{\tabcolsep}{5pt}
\begin{tabular}{lcccc}
\hline
Halo & $(v_0,v_{\rm esc})$ & Best-fit $(m_H,\delta)$
     & $\delta_{\rm pref}$ & $1\sigma$ slice \\
     & $[\mathrm{km\,s^{-1}}]$ & $[\mathrm{GeV},\mathrm{keV}]$
     & $[\mathrm{keV}]$ & $[\mathrm{keV}]$ \\
\hline
Lower speeds   & $(220,520)$ & $(1265,352)$ & $349$ & $340.7$--$360.5$ \\
Reference halo & $(238,544)$ & $(1080,369)$ & $369$ & $360.2$--$382.0$ \\
Higher speeds  & $(260,580)$ & $(660,384)$  & $400$ & $388.9$--$413.8$ \\
\hline
\end{tabular}
\caption{
Global best-fit points for the three halo assumptions.
The last two columns show the preferred mass splitting
and the intersection with the nominal $1\sigma$
profile-likelihood region at a fixed dark matter mass
of $m_H=1080\,\mathrm{GeV}$.
}
\label{tab:halo_profile_comparison}
\end{table}



\section{More Realistic Constraints}
\label{sec:solar_constraints}
The same $Z$-mediated inelastic scattering responsible for
the LZ recoil events can also lead to dark matter capture
in the Sun. Once captured, dark matter particles may
accumulate and annihilate, producing high-energy neutrinos
that can be constrained by IceCube. Previous studies of
TeV-scale Higgsino dark matter have shown that the strength
of this constraint depends on the subsequent evolution of
the captured particles. When elastic scattering allows
efficient thermalization, Pospelov and Ramani obtained
a lower bound of $\delta_m\gtrsim566\,\mathrm{keV}$,
excluding the Higgsino interpretation of the LZ event
under their assumptions~\cite{Pospelov:2026ewn}.
In contrast, Wu et al.\ showed that suppressed elastic
cooling can leave the captured particles on extended
orbits, reducing their annihilation rate and weakening
the IceCube bound to approximately $475\,\mathrm{keV}$
at 90\% C.L. They further found that the high-velocity
tail associated with the Large Magellanic Cloud (LMC)
can favor larger mass splittings in the LZ fit, allowing
a region consistent with the solar-neutrino constraint
~\cite{Wu:2026nhi}.

These studies are particularly relevant to the IDM,
where the two neutral inert scalars are also connected
by an off-diagonal $Z$ interaction. The solar capture
mechanism is therefore similar to that of Higgsino
dark matter, although the subsequent thermalization
and annihilation can differ between the two models.
Consequently, the IceCube bounds derived for Higgsino
dark matter cannot be directly applied to the IDM.
As shown in Sec.~\ref{sec:Galactic inputs}, the
preferred mass splitting in the IDM also depends
on the halo velocity distribution. We therefore
calculate solar capture and the subsequent evolution
of the captured IDM particles, before examining
how the LMC affects the compatibility between
the LZ interpretation and IceCube constraints.

\subsection{Solar capture and thermalization}
\label{sec:solar}
We begin with the LZ best-fit point in
Eq.~\eqref{eq:bestfit}, with $m_H=1080\,\mathrm{GeV}$
and $\delta=369\,\mathrm{keV}$, assuming that $H$
constitutes all of the dark matter in the reference halo.
As dark matter particles fall into the Sun, they gain
kinetic energy from the solar gravitational potential,
allowing them to scatter inelastically off heavy nuclei.
If their speed after scattering falls below the local
escape speed, they become gravitationally bound to the
Sun. Using the B16-GS98 solar model~\cite{Vinyoles:2016djt},
we obtain a capture rate of
\begin{equation}
  C_\odot \simeq 1.35\times10^{23}\,\mathrm{s^{-1}},
  \label{eq:solar_capture_rate}
\end{equation}
with iron providing the dominant contribution.


Once captured, IDM particles can continue to lose energy through elastic
scattering as inelastic upscattering becomes kinematically suppressed.
The original \texttt{micrOMEGAs} scan used tree-level elastic cross
sections. At this benchmark, the small Higgs-portal coupling gives a
spin-independent cross section of order $10^{-51}\,\mathrm{cm}^2$.
For the solar evolution, we therefore include the leading electroweak
corrections following Ref.~\cite{Abe:2015rja}, defining the portal
coupling at the reference Higgs momentum $q^2=m_h^2$. Combining the tree-level and loop amplitudes for the benchmark point in Eq.~\eqref{eq:bestfit}, we obtain
\begin{equation}
\sigma_{\mathrm{SI},p}\simeq1.0\times10^{-47}\,\mathrm{cm}^2,~~~~
\sigma_{\mathrm{SI},n}\simeq1.1\times10^{-47}\,\mathrm{cm}^2. 
\end{equation}


Having determined the scattering cross sections, we numerically
follow the orbital evolution of captured IDM particles gravitationally
captured by the Sun to investigate their subsequent energy loss and possible
escape. In this calculation, we track elastic scattering as well as two inelastic scattering
channels: endothermic scattering, $HN\to AN$, and exothermic
scattering, $AN\to HN$. For each channel, the scattering
probability is evaluated along the particle's orbit using the
local solar composition, density and nuclear temperature.
The two inelastic channels differ in their kinematics:
endothermic upscattering requires sufficient relative kinetic
energy to overcome the mass splitting, whereas exothermic
downscattering releases the excitation energy and can therefore
increase the particle's orbital energy. Whenever a collision
occurs, we determine the outgoing particle velocity using
two-body kinematics, taking into account the thermal motion
of the target nucleus. The resulting velocity is then used to
update the orbital energy and angular momentum and to determine
whether the particle remains gravitationally bound.

In addition to nuclear scattering, a particle in the excited
state $A$ can undergo the decay $A\to H\nu\bar{\nu}$.
To account for the competition between these two processes,
we sample the decay time from an exponential distribution
with the calculated mean lifetime and compare it with the
time until the next collision. If decay occurs first, the particle returns to the $H$ state,
with the excitation energy carried away by the neutrinos.
Since the recoil of the daughter particle is small, we neglect
it and retain the particle's orbital energy and angular momentum
at the decay step.

We find that, of the $128$ sampled trajectories, $125$ reach compact
orbits near the solar core, with a median time of
approximately $1.8\,\mathrm{Myr}$.
Once these compact orbits are reached, the estimated
thermal relaxation time is about $0.23\,\mathrm{Myr}$,
also much shorter than the age of the Sun.
These results support describing the retained dark matter
population by a thermal distribution near the solar core.
The evolution of the particle number $N_H$ is then
governed by
\begin{equation}
  \frac{dN_H}{dt}=C_{\rm ret}-C_A N_H^2,
  \qquad
  C_A=\frac{\langle\sigma v\rangle_{HH}}{V_{\rm eff}},
  \qquad
  \Gamma_{\rm ann}=\frac{1}{2}C_A N_H^2,
  \label{eq:solar_evolution}
\end{equation}
where $C_{\rm ret}\simeq(125/128)C_\odot$ is the effective
capture rate contributing to the thermalized population,
and $V_{\rm eff}$ denotes the effective volume of
the thermal core.

For this population, the characteristic timescale
for capture--annihilation equilibrium is approximately
$5.4\,\mathrm{Myr}$, much shorter than the age of the Sun.
The present annihilation rate is therefore
\begin{equation}
  \Gamma_{\rm ann}\simeq\frac{C_{\rm ret}}{2}
  \simeq6.6\times10^{22}\,\mathrm{s^{-1}}.
\end{equation}
The resulting annihilation into $W^+W^-$ and $ZZ$
final states produces a high-energy neutrino signal
that exceeds the adopted IceCube IC79 reference
for this spectrum by a factor of approximately
$145$~\cite{Baratella:2013fya,IceCube:2016yoy}.
Thus, the solar-neutrino constraint excludes the
original LZ best-fit point under the reference
halo assumptions.

\subsection{IceCube constrains and the high-speed tail}
\label{sec:solar_fast_window}

\begin{figure}[t]
  \centering
  \includegraphics[width=\linewidth]
  {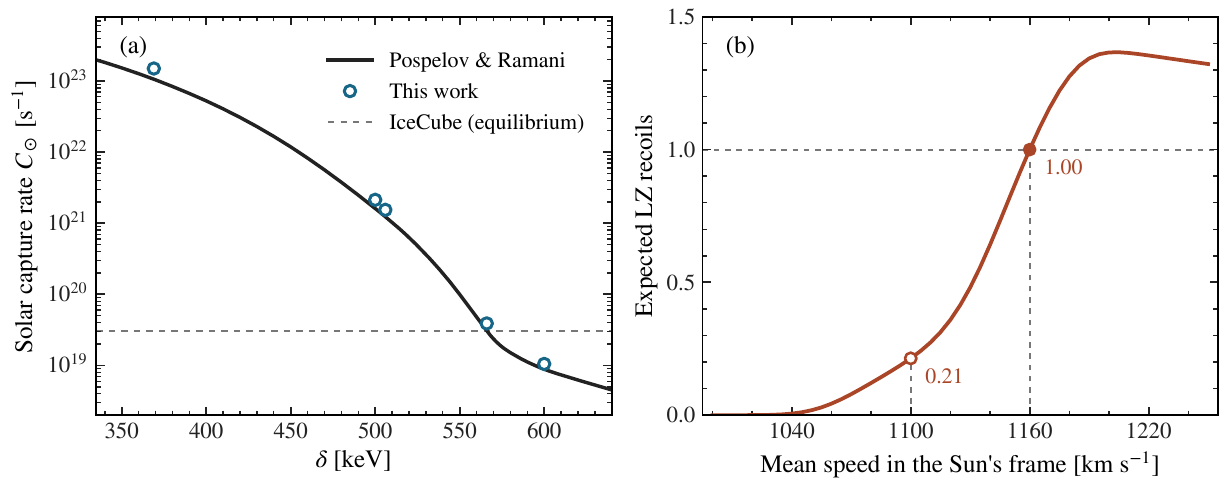}
  \caption{
  Solar capture and expected LZ recoil yield
  for a dark matter mass of $1.08\,\mathrm{TeV}$.
  Panel (a) compares our standard-halo capture
  calculation (blue circles) with the published
  Higgsino result (black curve). The gray dashed
  line indicates the capture rate corresponding
  to the IceCube $W^+W^-$ limit under
  capture--annihilation equilibrium.
  Panel (b) shows the expected number of selected
  inelastic LZ recoils at
  $\delta=635\,\mathrm{keV}$ as a function of
  the additional component's mean speed in
  the Sun's frame, with a fixed local density
  fraction $f_{\rm fast}=1.27\times10^{-4}$.
  The true recoil energies are restricted to
  $200$--$300\,\mathrm{keV}$, including the
  LZ detection efficiency for an exposure
  of $2.84\,\mathrm{tonne\,yr}$.
  }
  \label{fig:solar_capture_fast_tail}
\end{figure}

As shown in Sec.~\ref{sec:solar}, the original LZ
best-fit point in the IDM, with
$m_H=1080\,\mathrm{GeV}$ and
$\delta=369\,\mathrm{keV}$, is excluded by
the IceCube solar-neutrino constraint under
the reference halo assumptions, due to efficient
solar capture and subsequent annihilation.
To understand how this constraint depends on
the mass splitting, we first revisit the Higgsino
calculation of Pospelov and
Ramani~\cite{Pospelov:2026ewn}.
Using the same standard-halo assumptions and
inelastic scattering inputs, we independently
calculate the solar capture rate.
Figure~\ref{fig:solar_capture_fast_tail}(a)
compares our results (blue circles) with their
published capture curve (black line).
The gray dashed line indicates the capture rate
corresponding to the IceCube $W^+W^-$ limit
under capture--annihilation equilibrium
~\cite{IceCube:2025fcu}.
At $\delta=369\,\mathrm{keV}$, we obtain
$C_\odot\simeq1.5\times10^{23}\,\mathrm{s^{-1}}$,
compared with $1.1\times10^{23}\,\mathrm{s^{-1}}$
from the published result.
Using their capture curve and the two limiting
assumptions for elastic cooling, we recover
lower bounds on the Higgsino mass splitting
of approximately $507\,\mathrm{keV}$ and
$566\,\mathrm{keV}$ for negligible and
efficient cooling, respectively.

As illustrated in Fig.~\ref{fig:solar_capture_fast_tail}(a),
the solar capture rate decreases rapidly with
increasing mass splitting. Although a larger
splitting can therefore weaken the IceCube
constraint, it also raises the minimum dark matter
speed required to produce a high-energy recoil
at LZ. The possibility of satisfying both
constraints thus depends on the high-speed tail
of the local dark matter velocity distribution.
The LMC can enhance this
population, although the extent of the tail varies
among simulated Milky Way--LMC systems~\cite{Smith-Orlik:2023kyl,Zhang:2026hmg}.
In the published Auriga Halo~13 distribution,
the detector-frame speeds reach approximately
$950\,\mathrm{km\,s^{-1}}$.
Using this distribution directly, we find no
expected recoils in the $200$--$300\,\mathrm{keV}$
true-recoil range at $\delta=635\,\mathrm{keV}$.

Graham et al.\ considered an Auriga-inspired
high-speed tail extending to
$1100\,\mathrm{km\,s^{-1}}$ in the detector frame
and found Higgsino parameter regions compatible
with both LZ and IceCube~\cite{Graham:2026huo}.
A similar high-speed component has also been
considered for a lighter inert doublet
~\cite{Cerdeno:2026qio}.
Motivated by these studies, we consider an
additional high-speed population of the same
IDM dark matter particle $H$, sharing the
same mass and interactions as the bulk halo.
We identify a new IDM benchmark with
$m_H=1080\,\mathrm{GeV}$, $\delta=635\,\mathrm{keV}$,
$m_{H^\pm}-m_H\simeq8.33\,\mathrm{GeV}$,
$\lambda_L=-0.010$, and $\lambda_2=0.010$, yielding a thermal relic abundance of
$\Omega_Hh^2\simeq0.12$.
This benchmark can also account for the high-energy
LZ recoil while remaining consistent with the
IceCube results and constrains in Sec.\ref{sec:other_constraints}.

Figure~\ref{fig:solar_capture_fast_tail}(b)
shows the expected LZ recoil yield for this
benchmark as a function of the additional
component's mean speed in the Sun's frame.
We choose its fraction of the total local
dark matter density to be
$f_{\rm fast}=1.27\times10^{-4}$, giving
approximately one expected recoil at
$1160\,\mathrm{km\,s^{-1}}$, as indicated
by the filled circle.
Keeping this fraction fixed, reducing the
mean speed to $1100\,\mathrm{km\,s^{-1}}$
lowers the expected yield to $0.21$ events,
marked by the open circle.
For the benchmark at
$1160\,\mathrm{km\,s^{-1}}$, the corresponding
solar capture rate is
$1.7\times10^{19}\,\mathrm{s^{-1}}$, and
the predicted neutrino signal reaches only
about $56\%$ of the adopted IceCube reference
limit for the IDM annihilation spectrum.

Using the benchmark parameters given above,
we find destructive interference between the tree-level
and loop amplitudes. Including the leading electroweak correction
described in Sec.~\ref{sec:solar} reduces the spin-independent proton
cross section from $2.7\times10^{-48}\,\mathrm{cm}^2$ at tree level to
$2.4\times10^{-48}\,\mathrm{cm}^2$.
With this elastic cross section, the captured population thermalizes
efficiently, and capture and annihilation are nearly in balance.
The weaker neutrino signal therefore reflects the lower capture rate
at the larger mass splitting.
Thus, the additional high-speed population
allows the larger splitting to account for
the observed LZ recoil while reducing the
solar-neutrino signal.

\section{Conclusion}
\label{sec:conclusion}
The high recoil energy of the recent LZ event distinguishes it from the signals targeted by conventional low-energy elastic dark matter searches. Within the IDM, such an event can be naturally interpreted as an endothermic inelastic transition, $HN\to AN$, mediated by the off-diagonal $ZHA$ interaction. The elastic and inelastic scattering channels are governed by different interactions; consequently, suppressing the Higgs-mediated elastic scattering does not preclude the $Z$-mediated high-recoil signal.

Our analysis shows that the LZ likelihood favors a narrow
range of the neutral-scalar mass splitting, with a
profile-likelihood best-fit point at
$m_H=1080\,\mathrm{GeV}$ and
$m_A-m_H=369\,\mathrm{keV}$ under the reference
halo assumptions. This point satisfies the relevant
theoretical constraints and the bounds from collider
searches, the observed relic density, elastic direct
detection, and indirect detection. A particularly
interesting feature of the result is that the LZ
high-recoil event is much more sensitive to the
neutral-state splitting than to the overall dark
matter mass. We further find that the preferred
mass splitting depends on the halo velocity distribution,
reflecting the importance of its high-speed tail.

Solar-neutrino constraints further exclude the original
LZ best-fit point under the reference halo assumptions,
as the captured IDM particles thermalize efficiently
after including electroweak loop corrections.
Motivated by the LMC-enhanced high-speed tail, we
identify a new benchmark at $m_H=1080\,\mathrm{GeV}$
and $\delta=635\,\mathrm{keV}$ that can explain
the high-energy LZ recoil while remaining consistent
with the adopted IceCube limit.
The weaker neutrino signal arises primarily from
suppressed solar capture at the larger mass splitting,
rather than inefficient thermalization.
These results demonstrate the importance of the
halo velocity distribution and solar-neutrino
constraints for the IDM interpretation of the LZ event.


\appendix

\section{Validation of the Likelihood Method}
\label{app:lz_validation}

\begin{figure}[htbp]
  \centering
  \includegraphics[width=\linewidth]{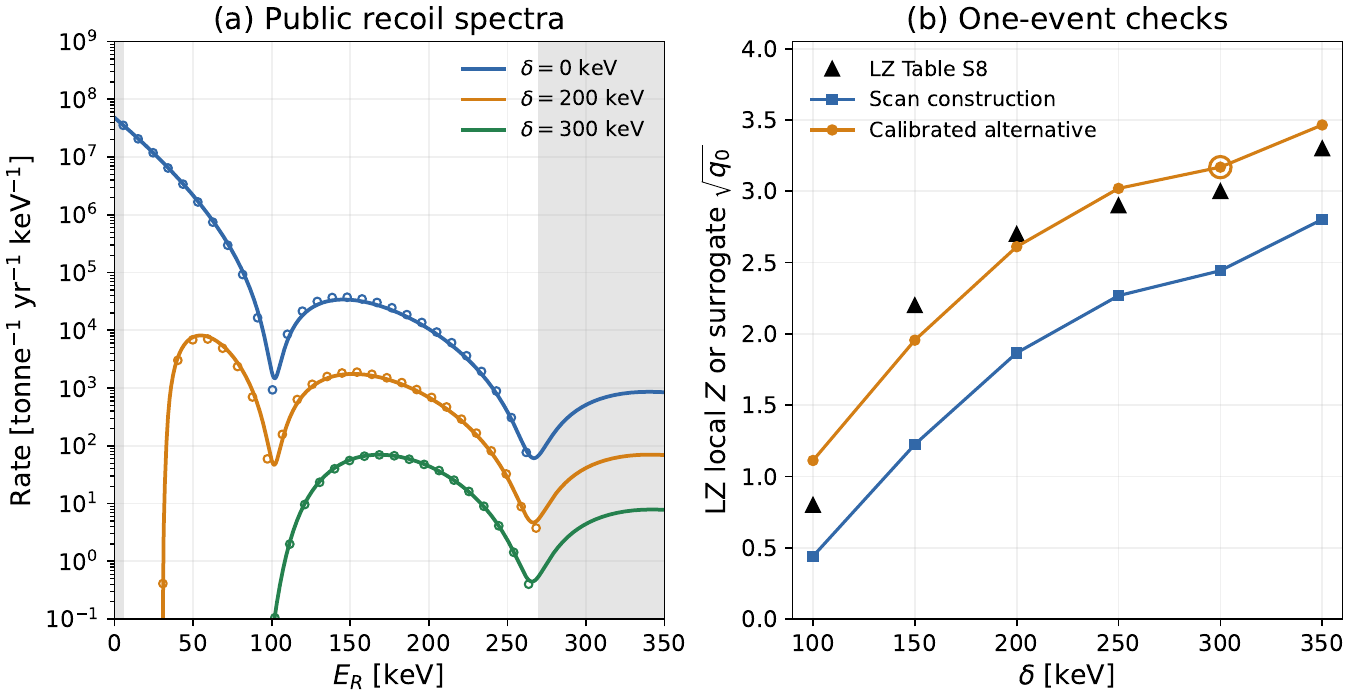}
  \caption{
  Comparison with public LZ benchmarks for
  $m_\chi=1\,\mathrm{TeV}$.
  Panel (a) shows the calculated recoil spectra (solid curves)
  and the published LZ results (open symbols).
  Panel (b) compares the published local significances with
  $\sqrt{q_0}$ obtained from our likelihood and a calibrated
  alternative. The circled point indicates the benchmark used
  for background calibration at zero energy displacement.
  }
  \label{fig:lz_validation}
\end{figure}

We first compare our recoil spectra with the public LZ results
for the isoscalar operator $\mathcal O_1^s$ in Fig.~1 of
Ref.~\cite{LZ:2026axp}. Using the same operator normalization
and halo parameters, we calculate the spectra for
$m_\chi=1\,\mathrm{TeV}$ and $\delta=0$, $200$, and
$300\,\mathrm{keV}$, as shown in the left panel of
Fig.~\ref{fig:lz_validation}. Without adjusting the overall
normalization, the root-mean-square differences in
$\log_{10}(R_{\rm calc}/R_{\rm LZ})$ are $0.067$, $0.084$,
and $0.030$ dex, respectively, over the common energy range
$5.4$--$269.9\,\mathrm{keV}$.

We then compare the likelihood results with the local
significances reported in Table~S8 of
Ref.~\cite{LZ:2026axp}, as shown in the right panel of
Fig.~\ref{fig:lz_validation}. Profiling over a nonnegative
signal strength, we define
$q_0=-2\ln(\mathcal L_0/\mathcal L_{\rm max})$,
where $\mathcal L_0$ is the profiled background-only
likelihood. At $\delta=300$ and $350\,\mathrm{keV}$,
we obtain $\sqrt{q_0}=2.44$ and $2.80$, compared with
the published values of $3.0$ and $3.3$, respectively.
To examine the dependence on the likelihood prescription,
we also consider the background calibration and energy
displacement adopted in Ref.~\cite{Wu:2026nhi},
using the $300\,\mathrm{keV}$ benchmark for calibration.
Despite the differences in local significance, both
prescriptions favor the same mass splitting,
$\delta=369\,\mathrm{keV}$, at fixed
$m_H=1080\,\mathrm{GeV}$ with the original IDM couplings
and reference halo.

\section*{Note added}
While this work was being completed, Ref.~\cite{Nomura:2026qyq} appeared, which considers a closely related scalar-doublet interpretation of the LZ high-recoil event. Our work provides a more detailed analysis within the IDM, systematically incorporating the dark matter observables and the relevant constraints, and performing a profile-likelihood analysis of the LZ event. This allows us to quantitatively determine the favored parameter region and identify the corresponding profile best fit point. 
In addition, we clarify that the high-recoil likelihood is primarily sensitive to the fine neutral state mass splitting rather than to the overall dark matter mass.

\acknowledgments

This work was supported by the National Natural Science Foundation of China (Grant Nos. 12605175 and 12605177), the Natural Science Foundation of Henan Province (Grant Nos. 27AQ140008), and the Shandong Provincial Natural Science Foundation (Grant No. ZR2023MA038).

\bibliographystyle{JHEP}
\bibliography{refs}

\end{document}